\RequirePackage[T1]{fontenc}
\documentclass[12pt]{article}

\usepackage[height=8.85in,width=6.45in]{geometry}

\usepackage[utf8]{inputenc}
\usepackage{amsmath}
\usepackage{amssymb}
\usepackage{mathtools}
\numberwithin{equation}{section}
\usepackage{slashed}
\usepackage{braket}
\usepackage[svgnames]{xcolor}
\usepackage[colorlinks,citecolor=DarkGreen,linkcolor=FireBrick]{hyperref}
\usepackage{cite}
\usepackage{graphicx}
\usepackage{tikz}
\usepackage{tikz-cd}
\usepackage{times}
\usepackage{courier}
\usepackage{bm}
\usepackage{subfig}
\usepackage{pifont}
\usepackage{nicematrix}
\usepackage{multirow}
\usepackage{colortbl}

\usepackage{xcolor}
\usepackage{mdframed}
\newenvironment{claim}{  \begin{mdframed}[linecolor=black!0,backgroundcolor=black!10]\noindent\itshape\ignorespaces}{\end{mdframed}}

\renewenvironment{figure}[1][]{
  \begin{originalfigure}[#1]
    \begin{mdframed}[linecolor=black!0,backgroundcolor=black!1]
}{
    \end{mdframed}
  \end{originalfigure}
}

\def\bR{\mathbb{R}}
\def\bZ{\mathbb{Z}}

\def\CL{{\mathcal L}}

\def\CN{{\mathcal N}}
\def\CO{{\mathcal O}}

\def\CR{{\mathcal R}}

\usepackage{datetime}
\usepackage{dsfont}

\begin{document}

\begin{titlepage}

\begin{flushright}
~
\end{flushright}

\vskip 3cm

\begin{center}

{\Large \bfseries Comments on compatibility between\vspace{5mm}\\
Conformal symmetry and Continuous higher-form symmetries}

\vskip 1cm
Yasunori Lee$^1$ and Yunqin Zheng$^{1,2}$
\vskip 1cm

\begin{tabular}{ll}
 $^1$&Kavli Institute for the Physics and Mathematics of the Universe, \\
 & University of Tokyo,  Kashiwa, Chiba 277-8583, Japan\\
 $^2$&Institute for Solid State Physics, \\
 &University of Tokyo,  Kashiwa, Chiba 277-8581, Japan\\
\end{tabular}

\vskip 1cm

\end{center}

\noindent

We study the compatibility between the conformal symmetry together with the unitarity and the continuous higher-form symmetries.
We show that the $d$-dimensional unitary conformal field theories are not consistent with
 continuous $p$-form symmetries for certain $(d,p)$, assuming that the corresponding conserved current is a conformal primary operator.
 We further discuss several dynamical applications of this constraint.

\end{titlepage}

\setcounter{tocdepth}{3}
\tableofcontents

\section{Introduction and Summary}

As quantum field theories (QFT) are subject to the renormalization group (RG) flow\cite{Wilson:1974mb},
one is naturally interested in its special points, one of which is the fixed point(s) at the end.
The theories at the fixed points are scale invariant by definition,
but they often enjoy even larger symmetry:
gapless theories are often conformal field theories (CFT),
and gapped theories are often topological quantum field theories (TQFT),
where the latter includes the trivially gapped theory with only one vacuum as a special case. 
To examine these fixed-point theories,
it is useful to find some universal features of the QFT that persist along the RG flow,
and it has been widely appreciated that the global symmetries and their 't Hooft anomalies are the two prominent ones.
In the simplest cases, the presence of nontrivial 't Hooft anomaly implies that
the IR fixed point cannot be a trivial theory with a symmetric ground state on an arbitrary spatial manifold\cite{LIEB1961407,Oshikawa_2000,Hastings_2004}.
However, it is natural to ask whether and how one can further constrain the fixed-point theories,
beyond just knowing whether they can be trivial or not.
Hopefully, the answer seems to be affirmative;
for \emph{gapped} IR fixed points,
it has been realized recently that certain systems with discrete global symmetries 
cannot flow to a symmetry-preserving TQFT in the IR \cite{Wan:2018djl, Cordova:2019bsd,Cordova:2019jqi}.

The aim of this short note is 
to find such more-refined constraints for \emph{gapless} fixed points,
focusing on continuous higher-form symmetries. 
As a main result, we find that
$d$-dimensional unitary CFTs are not compatible with continuous $p$-form symmetries
for some pairs of $(d,p)$. 
Some special instances have been discussed in the literature;
for $d=6$, supersymmetric CFTs are not compatible with continuous 1-form symmetry \cite{Cordova:2016emh},
and for $d=4$, CFTs are compatible with continuous 1-form symmetry only when it is chiral \cite{Argyres_1996}.
To the best of the authors' knowledge, the discussion for generic $(d,p)$ without imposing the supersymmetry has not been presented explicitly in the literature, and the purpose of this note is to fill this gap. 
The incompatibility between unitary CFTs and certain higher-form symmetries
can also be used to reinterpret some of the known results that
certain free field theories are scale invariant but not conformal \cite{Nakayama:2013is,ElShowk:2011gz}.

The paper is organized as follows.
In Section \ref{sec.confsymrep}, we review the conformal algebra and its representations.
In Section \ref{sec.unit}, we examine the scaling dimensions of the conserved currents for the continuous $p$-form symmetries,
and find that some of them violate the unitarity bounds and thus are forbidden in unitary CFTs.
We also apply the result to various examples and mention its consequences.
In Appendix \ref{app.SCFT}, we list additional constraints under the presence of supersymmetry.

\section{Conformal algebra and representation }
\label{sec.confsymrep}

To set up the notations, let us start by reviewing the conformal algebra and its representation.
Throughout, we work in the Euclidean space, with all-positive signature.
We follow the convention in \cite{Minwalla:1997ka}.

\paragraph{Conformal algebra}
The conformal symmetry algebra in $d$ spacetime dimensions is $\mathfrak{so}(d+1,1)$.
The generators are
the translations $P_\mu$,
the rotations $M_{\mu\nu}$,
the scaling $D$, and
the special conformal transformations $K_\mu$,
and they obey the following algebra 
\begin{equation}\label{ca}
	\renewcommand{\arraystretch}{1.2}
	\begin{array}{ccl}
		\left[M_{\mu\nu}, M_{\alpha\beta}\right] & = & -i\, (\eta_{\mu\beta} M_{\nu\alpha} + \eta_{\nu\alpha}M_{\mu\beta}- \eta_{\mu\alpha}M_{\nu\beta}-\eta_{\nu\beta}M_{\mu\alpha}),\\
      	\left[M_{\mu\nu}, P_{\alpha}\right] & = & -i\, (\eta_{\nu\alpha}P_{\mu}- \eta_{\mu\alpha}P_{\nu}),\\
      	\left[M_{\mu\nu}, D\right] & = & 0,\\
      	\left[M_{\mu\nu}, K_{\alpha}\right] & = & -i\, (\eta_{\nu\alpha}K_{\mu}- \eta_{\mu\alpha}K_{\nu}),\\
      	\left[D, P_{\mu}\right] & = & -i P_{\mu},\\
      	\left[D, K_{\mu}\right] & = & i K_{\mu},\\
      	\left[D, D\right] & = & 0,\\
      	\left[P_{\mu}, P_{\nu}\right] & = & 0,\\
      	\left[P_{\mu}, K_{\nu}\right] & = & -2i\, (\eta_{\mu\nu}D + M_{\mu\nu}),\\
      	\left[K_{\mu}, K_{\nu}\right] & = & 0,\\
	\end{array}
\end{equation}
where $\eta_{\mu\nu}$ is the metric with the Euclidean signature.

\paragraph{Representation of operators at the origin}
In a theory with the conformal symmetry,
the local fields $\CO$ can be organized into representations of the conformal algebra \eqref{ca}.
Under the radial quantization, there is a natural correspondence between
a local field $\CO(0)$ at the origin and
a state at the past infinity,
\begin{equation}
	\CO(0)
	\quad
	\leftrightarrow
	\quad
	\ket{\CO} \equiv \CO(0)\ket{0}	
\end{equation}
where $\ket{0}$ is the vacuum state. The Hermiticity condition for the conformal symmetry generators are 
\begin{equation}
	\renewcommand{\arraystretch}{1.2}
	\begin{array}{ccc}
		M_{\mu\nu}^\dagger & = & M_{\mu\nu},\\
		P_{\mu}^\dagger & = & K_{\mu},\\
		K_{\mu}^\dagger & = & P_{\mu},\\
		D^\dagger & = & -D.
	\end{array}
\end{equation}

To specify the representation,
one first needs to determine the Cartan subalgebra of \eqref{ca},
or equivalently, the maximal commuting subset of the generators.
A convenient choice for such subset is
$\{D, M_{12}, M_{34}, ..., M_{2m-1,2m}\}$ for $m=\lfloor\frac{d}{2}\rfloor$,
and then each operator $\CO$ (and hence each state $\ket{\CO}$) is labeled by
the scaling dimension $\Delta_\CO$ and
the highest weight\footnote{
	Here we adopt the orthogonal basis
	(rather than the fundamental-weight basis associated with the Dynkin labels),
	where $h_i \in \tfrac{1}{2}\bZ$ and $h_1 \geq \cdots \geq h_m$ for generic $d$,
	following \cite{Minwalla:1997ka}.
	(Note that for $d=4$, the representations will be labeled in terms of $\mathfrak{su}(2)\times \mathfrak{su}(2)$.)
	For more details of the representation theory including the relation between two bases,
	see standard textbooks e.g. \cite{FultonHarris}.
} $[\{h_i\}]_\CO$ of the $\mathfrak{so}(d)$ representation. 
The other generators are the ``ladder'' operators which raise or lower them.
For example, the momentum operator is represented in the standard way, 
\begin{equation}
    [P_{\mu},\CO(x)]\big|_{x=0} = -i \partial_\mu \CO(x)\big|_{x=0}
    \quad
    \leftrightarrow
    \quad
    P_\mu \ket{\CO} = -i \partial_{\mu} \ket{\CO},
\end{equation}
and therefore, $P_\mu|\CO\rangle$ has a scaling dimension $(\Delta_\CO+1)$.
In the following, we also restrict $\CO(0)$ to be a conformal primary operator,
which is defined to commute with the generators of special conformal transformations $K_\mu$,
and the corresponding state is annihilated by $K_\mu$,
\begin{equation}\label{Kmu}
    [K_\mu, \CO(0)]= 0
    \quad
    \leftrightarrow
    \quad
    K_{\mu} \ket{\CO}=0.
\end{equation}

\section{Constraint on unitary CFT from higher-form symmetries}
\label{sec.unit}

In this section, we will consider
the scaling dimension $\Delta_J$ and
the highest weights $[\{h_i\}]_J$ of the conserved currents associated with continuous higher-form symmetries,
and check whether they satisfy the unitarity bounds of the conformal representation.

\subsection{Unitarity bound in CFT}

As discussed in Section \ref{sec.confsymrep},
the state-operator correspondence maps
a conformal primary operator $\CO$
to
a conformal primary state $\ket{\CO}$,
and correspondingly, each state is labeled by
the scaling dimension $\Delta_\CO$
and the highest weights $[\{h_i\}]_\CO$. 
Here, the unitarity requires all states to have non-negative norms,
i.e.~$\langle \CO|\CO\rangle\geq 0$ in the Euclidean spacetime.
For the descendant state $\prod_{i=1}^n P_{\nu_i} \ket{\CO}$, this can be rephrased as
all the eigenvalues of the matrix 
\begin{equation}\label{AKP}
    \bra{\CO}
    \prod_{j=1}^n K_{\mu_j}
    \prod_{i=1}^n P_{\nu_i}
    \ket{\CO}
\end{equation}
should be non-negative.
The constraint for $n=1$ has been completely solved in \cite{Mack:1975je} for $d=4$,
and in \cite{Minwalla:1997ka} for general dimensions, which we summarize in Table~\ref{tab:UB}.
It turns out that there are no further constraints coming from $n\geq 2$ unless $\CO$ is a Lorentz scalar (i.e.~$\mathfrak{so}(d)$ singlet)
\cite{Mack:1975je}.

\begin{table*}
	\begin{align*}
	    \renewcommand{\arraystretch}{1.2}
    	\begin{array}{|c|c|c|cl| }
        	\hline
        	d
        	& \text{Lorentz Algebra}
        	& \text{Representation}
        	& \multicolumn{2}{c|}{\text{Unitarity Bound}~~~ \Delta_\CO\geq }\\
        	\hline
        	\multirow{3}{*}{$3$} & \multirow{3}{*}{$\mathfrak{so}(3)$} & \multirow{3}{*}{$[h]_{\CO}$}
        		 & 0 	& (h=0)\\
        	&&& 1 	  & (h=\frac{1}{2})\\
        	&&& h+1 & (h\geq 1)\\
        	\hline
        	\multirow{4}{*}{$4$} & \multirow{4}{*}{$\mathfrak{so}(4)$} & \multirow{4}{*}{$[h_1, h_2]_{\CO}$}
        		 & 0 			  & (h_1=h_2=0)\\
        	&&& h_1+1 		 & (h_1>0, h_2=0)\\
        	&&& h_2+1 		 & (h_1=0, h_2>0)\\
        	&&& h_1+h_2+2 & (h_1>0, h_2>0)\\
        	\hline
        	\multirow{4}{*}{$5$} & \multirow{4}{*}{$\mathfrak{so}(5)$} & \multirow{4}{*}{$[h_1, h_2]_{\CO}$}
        		 & 0 		& (h_1=h_2=0)\\
        	&&& 2 		 & (h_1=h_2=\frac{1}{2})\\
        	&&& h_1+2 & (h_1=h_2\neq 0, \frac{1}{2})\\
        	&&& h_1+3 & (h_1>h_2)\\
        	\hline
        	\multirow{4}{*}{$6$} & \multirow{4}{*}{$\mathfrak{so}(6)$} & \multirow{4}{*}{$[h_1, h_2, h_3]_{\CO}$}
        		 & 0 		& (h_1=h_2=h_3=0)\\
        	&&& h_1+2 & (h_1=h_2=|h_3|\neq 0)\\
        	&&& h_1+3 & (h_1=h_2>|h_3|)\\
        	&&& h_1+4 & (h_1>h_2)\\
        	\hline
    	\end{array}
	\end{align*}
	\caption{
    	Unitarity bounds of local primary operators
    	from the non-negativity of the norms of their first descendants. 
    	The highest weights $h_i$'s are all in terms of the orthogonal basis.
    }\label{tab:UB}
\end{table*}

\subsection{Unitarity bounds for conserved currents of higher-form symmetry}
\label{sec.UBforpformsymmetry}

A continuous $p$-form global symmetry is accompanied by
a $(p+1)$-form current $J=J^{\mu_{1} \cdots \mu_{p+1}} dx^{\mu_1}\wedge...\wedge dx^{\mu_{p+1}}$ which is conserved
\begin{equation}\label{cons}
    d*J=\partial_{\mu_1} J^{\mu_{1} \cdots \mu_{p+1}}=0,
\end{equation}
and the $p+1$ indices are fully anti-symmetrized~\cite{Gaiotto:2014kfa}.
The conserved charge is defined by integrating $\ast J$ over a $(d-p-1)$-dimensional submanifold $\Sigma_{d-p-1}$
\begin{equation}
    Q= \int_{\Sigma_{d-p-1}} *J \ ,
\end{equation}
and the charged operators are supported on $p$-cycles linked by $\Sigma_{d-p-1}$.
The background field for the $p$-form symmetry is $(p+1)$-form $B^{(p+1)}$, which couples to the action via the coupling term 
\begin{equation}
    \int B^{(p+1)}*J\ .
\end{equation}
This term is invariant under the background gauge transformation $B^{(p+1)}\to B^{(p+1)}+ d\lambda^{(p)}$, due to the conservation condition \eqref{cons}.

\begin{table*}
	\begin{align*}
	    \renewcommand{\arraystretch}{1.2}
    	\begin{array}{|l|cc|cc|cc|cc|}
        	\hline
        	& \multicolumn{2}{c|}{d=3} 
        	& \multicolumn{2}{c|}{d=4}
        	& \multicolumn{2}{c|}{d=5}
        	& \multicolumn{2}{c|}{d=6}\\
        	\cline{2-9}
        	& \Delta_J & [h]_J
        	& \Delta_J & [h_1, h_2]_J
        	& \Delta_J & [h_1, h_2]_J
        	& \Delta_J & [h_1, h_2, h_3]_J\\
        	\hline
        	p=0, J^{\mu}
        	& 2 &  [1]
        	& 3 & [\frac{1}{2}, \frac{1}{2}]
        	& 4 & [1,0]
        	& 5 & [1,0,0] \\
        	p=1, J^{\mu\nu}
        	& 1 & [1]
        	& 2 & [1,0]\oplus [0,1]
        	& 3 & [1,1]
        	& 4 & [1,1,0]\\
        	p=2, J^{\mu\nu\rho}
        	&   &
        	& 1 & [\frac{1}{2}, \frac{1}{2}]
        	& 2 & [1,1]
        	& 3 & [1,1,1]\oplus [1,1,-1]\\
        	p=3, J^{\mu\nu\rho\sigma}
        	&   &
        	&   &
        	& 1 & [1,0]
        	& 2 & [1,1,0]\\
        	p=4, J^{\mu\nu\rho\sigma\eta}
        	&   &
        	&   &
        	&   &
        	& 1 & [1,0,0]\\
        	\hline
    	\end{array}
	\end{align*}
	\caption{
	    Scaling dimensions and $\mathfrak{so}(d)$ Lorentz representations 
	    of the conserved currents of $p$-form symmetry in $d=3,4,5,6$.  
	}\label{tab:QN}
\end{table*}

We will assume that, when a CFT has continuous $p$-form symmetry $G$,\footnote{
	Since $p$-form symmetries with $p\geq 1$ must be Abelian,
	$G$ is either $U(1)$ or $\bR$ (or multiple copies thereof).
}
there is at least one operator charged under it.
This implies that the conserved current should not be a derivative of another operator,
since otherwise the charge $Q$ vanishes,
meaning that there is no charged operator in the theory and hence the symmetry is decoupled 
\cite{Cordova:2018cvg}.
In other words, the current should be a conformal primary operator.
Furthermore, by applying the state-operator correspondence to the conservation condition \eqref{cons},
one finds that the first descendant of the primary state $\ket{J}$ is a null state.
Hence $\ket{J}$ belongs to a short conformal multiplet and the unitarity bound tabulated in Table~\ref{tab:UB} must be saturated. 

\begin{table*}
	\begin{align*}
	    \renewcommand{\arraystretch}{1.2}
    	\begin{array}{|c|c|c|c|c|}
        	\hline
        	& d=3
        	& d=4
        	& d=5
        	& d=6 \\
        	\hline
        	p=0
        	& \text{\textcolor{DarkGreen}{\ding{51}}}
        	& \text{\textcolor{DarkGreen}{\ding{51}}}
        	& \text{\textcolor{DarkGreen}{\ding{51}}}
        	& \text{\textcolor{DarkGreen}{\ding{51}}}\\
        	p=1
        	& \text{\textcolor{red}{\ding{55}}}
        	& \text{\textcolor{DarkGreen}{\ding{51}}}: \text{if chiral}
        	\quad \text{\textcolor{red}{\ding{55}}}: \text{otherwise}
        	& \text{\textcolor{DarkGreen}{\ding{51}}}
        	& \text{\textcolor{DarkGreen}{\ding{51}}}\\
        	p=2
        	&
        	& \text{\textcolor{red}{\ding{55}}}
        	& \text{\textcolor{red}{\ding{55}}}
        	& \text{\textcolor{DarkGreen}{\ding{51}}}: \text{if chiral}
        	\quad \text{\textcolor{red}{\ding{55}}}: \text{otherwise}\\
        	p=3
        	&
        	&
        	& \text{\textcolor{red}{\ding{55}}}
        	& \text{\textcolor{red}{\ding{55}}}\\
        	p=4
        	&
        	&
        	&
        	& \text{\textcolor{red}{\ding{55}}}\\
        	\hline
    	\end{array}
	\end{align*}
	\caption{
    	$p$-form symmetries that saturates (\textcolor{DarkGreen}{\ding{51}}) or violates (\textcolor{red}{\ding{55}}) the unitarity bound, in spacetime dimensions $d=3,4,5,6$.
    	``Chiral'' means both the self-dual and anti-self-dual components of the current are conserved, i.e.~$d* J=0$ and $dJ=0$.
	}\label{tab:MR}
\end{table*}

Note that the symmetry charge $Q$ by definition has to commute with the conformal algebra.
In particular, the scaling dimension of $Q$ has to vanish $\Delta_Q=0$,
and this fixes the scaling dimension of the corresponding current to be $\Delta_J= d-p-1$.
In Table~\ref{tab:QN}, we enumerate them along with the $\mathfrak{so}(d)$ representation labeled by the highest weights.
By comparing it with the unitarity bounds in Table~\ref{tab:UB},
we can determine the pairs $(d,p)$ which are not compatible with unitary CFT,
as summarized in Table~\ref{tab:MR}.
For instance, when $d=3$, the current for the 1-form symmetry violates the unitarity bound since $\Delta_J=1< h+1=2$. We will refer to the symmetry whose conserved current violates  the unitarity bound as the ``forbidden'' symmetry. We summarize our main result as a theorem:
\begin{claim}
\textbf{Theorem}: A unitary CFT cannot have the ``forbidden'' $p$-form symmetry (\textcolor{red}{\ding{55}}) whose conserved current is the conformal primary operator. 
\end{claim}

The cases $(d,p)= (4,1)$ and $(6,2)$ deserve additional comments;
in these two cases, the conserved currents are in reducible representations of Lorentz symmetry,
and can be decomposed into self-dual (SD) and anti-self-dual (ASD) currents, each of which is irreducible.
For instance, for $(d,p)=(4,1)$,
the SD and ASD currents are defined as
$J^{\mu\nu}_{\pm}= \frac{1}{2}(J^{\mu\nu} \pm \frac{1}{2}\epsilon^{\mu\nu\rho\sigma}J_{\rho\sigma})$,
belonging to irreducible representations $[1,0]$ and $[0,1]$ respectively. Their scaling dimensions are identical, i.e.~$\Delta_J=\Delta_{J_\pm}=2$. From the unitarity bound in table \ref{tab:UB}, it follows that in a unitary CFT, both the SD and ASD currents should be conserved, $\partial_\mu J^{\mu}_{\pm}=0$. We will call such global symmetries and their currents ``chiral". On the contrary, a unitary CFT in $4d$ is not compatible with non-chiral 1-form global symmetries. Similar comments apply to $(d,p)=(6,2)$.

\subsection{Dynamical applications}
\label{sec.da}

The conflicts between the conformal symmetry (together with the unitarity) and higher-form symmetries
lead to several dynamical consequences.
Starting from a Lorentz-invariant $d$-dimensional QFT
with a ``forbidden'' continuous $p$-form global symmetry $G$
at arbitrary energy scale,
we would like to ask 
\begin{enumerate}
    \item[Q1.] UV completion: if there exists a UV fixed point
    which flows to the original QFT by turning on certain $G$-symmetric relevant operator,
    can the UV fixed point be a unitary $G$-symmetric  CFT?
    \item[Q2.] IR fate: if we turn on a $G$-symmetric relevant coupling and flow down along the RG,
    what will be the IR fixed point? Can it be a unitary $G$-symmetric  CFT?
\end{enumerate}
Below, we propose that there can be following dynamical scenarios; 
namely, the UV or IR fixed point theory can be
\begin{enumerate}
 \item[\hypertarget{scenario1}{1.}] a unitary CFT, but the $p$-form symmetry $G$ is decoupled. 
 \item[\hypertarget{scenario2}{2.}] scale invariant but not conformal, and the $p$-form symmetry $G$ may or may not decouple. 
 \item[\hypertarget{scenario3}{3.}] non-unitary. 
 \item[\hypertarget{scenario4}{4.}] gapped TQFT (including a trivial theory). 
\end{enumerate}
\begin{table}[]
    \centering
		\begin{tabular}{|r|c|c|c|c|}
		    \hline
		    & \multicolumn{3}{c|}{gapless} & gapped\\
		    \cline{2-5}
			& \multicolumn{2}{c|}{unitary} & non-unitary & \multirow{4}{*}{4}\\
			\cline{2-4}
			& conformal & scale inv. & \multirow{3}{*}{3} & \\
			\cline{1-3}
			symmetry not decoupled & \cellcolor{lightgray} & \multirow{2}{*}{2} &&\\
			\cline{1-2}
			decoupled &  1 &&&\\
			\hline
		\end{tabular}
    \caption{
    	Summary of dynamical scenarios in the presence of forbidden (\textcolor{red}{\ding{55}}) symmetry.
    }
    \label{tab:summaryscenarios}
\end{table}
See Table \ref{tab:summaryscenarios} for a comparison between different scenarios. Some comments are in order. 
\begin{itemize}
    \item The \hyperlink{scenario1}{scenario 1} is consistent from the unitarity bound analyses in Section~\ref{sec.UBforpformsymmetry},
    because there we assumed that the current does not decouple.

    \item The \hyperlink{scenario2}{scenario 2} appears somewhat exotic,
    since the scale invariance usually comes together with the conformal invariance,
    which was shown to be always the case in $d=2$~\cite{Polchinski:1987dy}.
    However, in higher dimensions, there is no such proof, 
    and in fact there are some counterexamples \cite{ElShowk:2011gz,Komargodski:2014lecture,Nakayama:2013is}.
    Below, we will clarify that those scale-invariant but non-conformal theories actually possess ``forbidden'' symmetries,
    which do not allow them to be unitary and conformal at the same time.
    
    \item Regarding the \hyperlink{scenario4}{scenario 4},
    	the UV fixed point theory is unlikely to be a TQFT with a unique vacuum, 
    because the only local operator would be the identity operator,
    and hence  there is no relevant operator to trigger the RG flow to the original theory.
    If there are multiple vacua, one can analyze the TQFT in one particular vacuum, and the same conclusion follows. 
    
    \item Although TQFT is a special case of CFT,
    it is actually compatible with continuous $p$-form symmetries
    forbidden by the unitarity bound.
    This is because the current annihilates the vacuum $J|0\rangle=0$,
    and therefore $\langle 0|J^\dagger D J|0\rangle=0$, where we do not find the nontrivial inequality following from \eqref{AKP}. 
\end{itemize}
In the following, we will look into various concrete examples
and see which scenario takes place in each case.

\subsubsection{Free compact scalar}
\label{sec.freeboson}

Consider a free real compact scalar $\phi$ with $\phi\sim \phi+2\pi R$ in $d$ spacetime dimensions.
The Lagrangian is given by
\begin{eqnarray}\label{freescalar}
    \CL=\frac{1}{2}(\partial_\mu \phi)^2,
\end{eqnarray}
and $\phi$ (and hence the radius $R$) has the scaling dimension $\frac{d-2}{2}$.
This theory enjoys two global symmetries ($G^{(p)}$ denotes the $p$-form symmetry group $G$):
\begin{itemize}
	\item $U(1)^{(0)}$ electric shift symmetry:
	$\phi\to \phi+\lambda$.  
	\item $U(1)^{(d-2)}$ topological symmetry:
	the conserved current is $J^{\mu_1 \cdots \mu_{d-1}}= \frac{1}{2\pi R} \epsilon^{\mu_1\cdots \mu_{d-1}\mu_{d}}\partial_{\mu_d} \phi$, where we have normalized the current so that the charge is an integer. 
\end{itemize}
Let us analyze the RG flow of this theory.
For $d=2$, the theory is scale invariant, and is also conformal, known as the free boson CFT.
For $d\geq 3$, the radius $R$ is dimensionful with a positive scaling dimension, and hence it grows under the RG flow.

\paragraph{UV fixed point ($\bm{R\to 0}$):}
The radius of $\phi$ shrinks to zero, and the electric shift symmetry $U(1)^{(0)}$ becomes trivial and vanishes.
However, since the scalar is still compact, the topological symmetry survive, although it becomes $\bR^{(d-2)}$ symmetry.
To see how $U(1)$ becomes $\bR$ and how it acts on the gapless degrees of freedom,
it is convenient to perform an $S$-duality transformation.
Under the duality, $d\phi$ is mapped to $-\frac{1}{2\pi} (\ast F)$, where $F=dA$ for  a $(d-2)$-form $A$.
The Lagrangian \eqref{freescalar} then becomes
\begin{equation}
	\label{dualtheory}
	\CL=-\frac{1}{8\pi^2}F \wedge \ast F
\end{equation} 
where $A$ has the scaling dimension $\Delta_A=\frac{d-2}{2}$. 
Note that for $d=3$, \eqref{dualtheory} is the standard free Maxwell theory with $A$ being a 1-form gauge field.
The periodicity of the $(d-2)$-form gauge field is $A\sim A+ \frac{2\pi \eta}{R}$,
where $\eta$ is a $(d-2)$-form flat gauge field with the scaling dimension $d-2$, and integrates to 1 on $S^{d-2}$.

Under the duality, the $U(1)^{(d-2)}$ topological symmetry of the original theory \eqref{freescalar} becomes the $U(1)^{(d-2)}$ electric shift symmetry of the dual theory \eqref{dualtheory}.
At the UV fixed point, the compact $U(1)$ gauge field becomes non-compact, i.e.~an $\bR$ gauge field.
Since the $\bR^{(d-2)}$ symmetry acts on the gapless gauge field $A$, it does not decouple,
and the general result on the unitarity bound in Section~\ref{sec.UBforpformsymmetry} applies;
from Table~\ref{tab:MR}, we find that the $(d-2)$-form symmetry is forbidden for all the dimensions $d=3,4,5,6$,
and 
we conclude that the UV fixed point cannot be a unitary CFT (i.e.~the \hyperlink{scenario1}{scenario 1} is ruled out).
Our result is consistent with the analyses in \cite{Jackiw:2011vz, Nakayama:2013is, ElShowk:2011gz, Komargodski:2014lecture}
where it has been explicitly shown (by computing the correlation function and conformal currents) that
the theory \eqref{freescalar} is unitary, scale invariant but not conformal (i.e.~the \hyperlink{scenario2}{scenario 2} is realized). 

\paragraph{IR fixed point ($\bm{R\to \infty}$): }
The radius of $\phi$ becomes infinite, and hence one has a theory of a non-compact real free scalar.
The $U(1)^{(0)}$ electric shift symmetry becomes a $\bR^{(0)}$ shift symmetry,
while the $U(1)^{(d-2)}$ topological symmetry disappears because the current $J$ becomes trivial.
Therefore, the IR theory does not have any higher-form symmetry,
and the unitarity bounds do not provide any further obstructions to conformal invariance.
As shown explicitly in \cite{Jackiw:2011vz, Nakayama:2013is, ElShowk:2011gz, Komargodski:2014lecture},
the non-compact free scalar theory is a CFT on a flat manifold in any dimensions, which is consistent with our result.
On a curved manifold, one needs to add a term $\int d^dx \CR \phi^2$ proportional to the scalar curvature $\CR$
to make the conformal symmetry manifest.
Note that this coupling is forbidden when $\phi$ is compact, because it is not invariant under the shift $\phi\to \phi+2\pi R$.

\subsubsection{Free Maxwell theory}
\label{sec.maxwelltheory}

Consider another class of free field theory - the free 	Maxwell theory. The Lagrangian is
\begin{eqnarray}\label{freemaxwell}
    \CL= -\frac{1}{8\pi^2} F\wedge * F
\end{eqnarray}
where $F=dA$ for a compact $U(1)$ $1$-form gauge field $A$.
The periodicity is $A\sim A+ \frac{2\pi \eta}{R}$, where $\eta$ is a $1$-form flat gauge field which integrates to 1 on $S^1$,
and $R$ is a dimensionful scalar specifying the (inverse) radius of $A$.
The theory again enjoys two global symmetries:
\begin{itemize}
    \item $U(1)^{(1)}$ electric shift symmetry: $A\to A+\xi$, where $\xi$ is a $1$-form flat gauge field. 
    \item $U(1)^{(d-3)}$ topological symmetry: the conserved current is $J^{\mu_1\cdots \mu_{d-2}}= \frac{R}{4\pi}\epsilon^{\mu_1\cdots \mu_{d-1}\mu_{d}}F_{\mu_{d-1}\mu_{d}}$, where we have normalized the current so that the charge is an integer. 
\end{itemize}
Let us analyze the RG flow.
The only dimensionful coupling constant is $R$ with scaling dimension $\Delta_R=\frac{4-d}{2}$.
When $d=3$, $\Delta_R>0$ and $R$ increases under the RG flow;
as discussed in Section~\ref{sec.freeboson}, the $3d$ free Maxwell theory is dual to free compact scalar theory,
and one can directly refer to the results there.
When $d=4$, $\Delta_R=0$ and $R$ is invariant under the RG flow;
the theory \eqref{freemaxwell} is not only scale invariant, but is also conformal, known as the Maxwell CFT.
Finally, when $d\geq 5$, $\Delta_R<0$ and $R$ decreases under the RG flow.
Let us take a closer look at the fixed points. 

\paragraph{UV fixed point ($\bm{R\to \infty}$):}
The $U(1)^{(1)}$ electric shift symmetry disappears.
By performing the $S$-duality as in Section~\ref{sec.freeboson},
one can see that the $U(1)^{(d-3)}$ topological symmetry becomes $\bR^{(d-3)}$,
and acts as a electric shift symmetry on the gapless $(d-3)$-form dual gauge field.
Note that for $d=5,6$, by comparing with the results in Table~\ref{tab:MR}, the conserved currents violate the unitarity bound.
Hence the UV fixed point cannot be a unitary CFT. 
Our result is again consistent with \cite{Jackiw:2011vz, Nakayama:2013is, ElShowk:2011gz, Komargodski:2014lecture},
where the UV fixed point has been proved to be 
scale invariant but not conformal. 

\paragraph{IR fixed point ($\bm{R\to 0}$):} 
The $U(1)^{(1)}$ electric symmetry becomes $\bR^{(1)}$, while $U(1)^{(d-3)}$ decouples.
The $\bR^{(1)}$ symmetry is compatible with the unitarity bound for $d\geq 5$,
and we cannot rule out the possibility of unitary CFT.

\subsubsection{Four-derivative Maxwell theory in $6d$}
\label{sec.fourderivative}

In Section \ref{sec.maxwelltheory}, we have noted that the $4d$ free Maxwell theory \eqref{freemaxwell} is a CFT.
In $6d$, it has been shown \cite{Giombi:2015haa, Tseytlin:2013fca} that the \emph{four-derivative} Maxwell theory
\begin{eqnarray}\label{fourderivativeMT}
\CL= \frac{1}{4e^2} G_{\mu\nu} \nabla^2 G^{\mu\nu}
\end{eqnarray} 
is also conformal.
Here, $G=dB$ is the field strength of the 1-form gauge field $B$,
and $e$ is the coupling constant.
As always, this $6d$ theory has two global symmetries: 
\begin{itemize}
	\item $U(1)^{(1)}$ electric shift symmetry: $B\to B+\xi$, where $\xi$ is a 1-form flat gauge field. 
	\item $U(1)^{(3)}$ topological symmetry: the conserved current is $J= \frac{1}{2\pi}*dB$.
\end{itemize}
Both symmetries act nontrivially on the gapless modes.
For the former, it is obvious because it shifts the gauge field $B$.
For the latter, it is easier to consider the dual theory as in the previous examples;
the dual theory is also a free theory, with the quadratic Lagrangian $F^{(4)} \wedge \nabla^{-2} * F^{(4)}$,
where $F^{(4)}=dC^{(3)}$ is the 4-form field strength, 
and the 3-form global symmetry is just the shift symmetry $C^{(3)}\to C^{(3)}+ \eta^{(3)}$.

Recalling that the 3-form symmetry $U(1)^{(3)}$ is not compatible with the conformal symmetry and the unitarity,
the theory can only be a non-unitary CFT.
Indeed, it has been explicitly computed in \cite{Giombi:2016fct} that
the coefficient of the two point correlation function of the stress-energy tensors in the theory \eqref{fourderivativeMT} is negative,
which implies that there exist negative-norm states and thus the theory is indeed non-unitary.

\subsubsection{QED$_4$ }

We now proceed to discuss an interacting theory, QED$_4$.
The Lagrangian is
\begin{eqnarray}
\CL= -\frac{1}{4e^2} F\wedge\ast F + \sum_{i=1}^{N_f} \bar\psi_i (\slashed \partial - i \slashed A) \psi_i.
\end{eqnarray}
The only higher-form global symmetry is the $U(1)^{(1)}$ topological symmetry, with the current $J= \frac{1}{2\pi}*dA$.
Note that this symmetry is non-chiral unless the gauge field decouples from the fermion.

It is well-known that the 1-loop beta function of QED$_4$ is positive for arbitrary $N_f$.
Hence in the IR, the theory becomes free and the fermion and gauge sectors decouple.
In the low energy limit, the theory flows to a unitary CFT because both
the free Maxwell theory in $4d$, as well as the free fermion, are  unitary CFTs.
This is consistent with our main result in Section \ref{sec.UBforpformsymmetry}. 

In the UV limit, QED$_4$ is regarded as ill-defined,
because there is a Landau pole where the coupling constant diverges.
Therefore, it should only be regarded as a low-energy effective theory.
One way to make it well-defined in the UV limit is
to embed it into a larger theory (possibly a non-Abelian gauge theory).
Suppose there exists such a parent theory, and the embedding preserves the non-chiral $U(1)^{(1)}$ symmetry in the QED$_4$.
Then, our result might be able to constrain the property of the UV fixed point of the parent theory.
Similar analysis can also be applied to $d \geq 5$.

\subsubsection{A variant of QED$_6$}

Our last example is a variant of QED$_6$, which was discussed in \cite{Giombi:2015haa}.
The Lagrangian is 
\begin{eqnarray}\label{FDF}
\CL= \frac{1}{4e^2} F_{\mu\nu} \nabla^2 F^{\mu\nu} + \sum_{i=1}^{N_f} \bar\psi_i (\slashed \partial - i \slashed A)\psi_i
\end{eqnarray}
where the ordinary Maxwell term is replaced by the four-derivative Maxwell term.
As discussed in Section~\ref{sec.fourderivative}, there are two global symmetries,
but only the  $U(1)^{(3)}$ symmetry survives once the gauge field couples to the fermion with charge 1.
Since this symmetry is not compatible with the conformal invariance and the unitarity,
it is tempting to use it to constrain the RG flow of \eqref{FDF}.

It was computed in \cite{Giombi:2015haa} that the beta function of \eqref{FDF} is negative,
$\beta_e= -\frac{\epsilon}{2}e- \frac{N_f}{120\pi^3} e^3 + \CO(e^5)$,
where $\epsilon=6-d$ is the parameter introduced in the dimensional regularization.
This means that in the UV limit, the gauge sector decouples from the fermion,
and hence the UV fixed point is a non-unitary CFT analyzed in Section \ref{sec.fourderivative},
tensored with a unitary free fermion CFT.
In the IR, the theory \eqref{FDF} becomes strongly-coupled.

If the RG flow does not explicitly break the $U(1)^{(3)}$ symmetry,
then our main theorem can be used to constrain the possible scenario of the IR fixed point.
Note that although the pure four-derivative Maxwell term is non-unitary,
the IR fixed point may still be unitary, when $N_f$ is sufficiently large.

\section*{Acknowledgements}

We thank Yu Nakayama, Kantaro Ohmori, Masaki Oshikawa, Yuji Tachikawa, Juven Wang, Masahito Yamazaki and Kazuya Yonekura for helpful correspondences. We especially thank Yuji Tachikawa for detailed comments on the draft. 
Y.L. is partially supported by the Programs for Leading Graduate Schools, MEXT, Japan, via the Leading Graduate Course for Frontiers of Mathematical Sciences and Physics,
and also by JSPS Research Fellowship for Young Scientists.
Y.Z. is partially supported by WPI Initiative, MEXT, Japan at IPMU, the University of Tokyo.

\newpage

\appendix

\section{Unitarity bounds for conserved currents in SCFT}
\label{app.SCFT}

Here, we show the constraints from the unitarity bounds for conserved currents associated with the higher-form symmetries
in superconformal field theories (SCFT).
We assume that the higher-form symmetries commute with the superconformal algebra, and thus the currents do not carry $R$-charges.
Due to the additional generators $Q_\alpha$ and $S_\alpha$ compared to the ordinary conformal algebra,
there are additional states, and correspondingly all the eigenvalues of the matrix 
\begin{equation}\label{AKPS}
    \bra{\CO}
    \prod_{j=1}^n Q_{\alpha_j}
    \prod_{i=1}^n S_{\beta_i}
    \ket{\CO}
\end{equation}
should also be non-negative.  
This gives rise to additional unitarity bounds~\cite{Cordova:2016emh, Minwalla:1997ka},
and as a result, continuous $1$-form symmetries in $d=5,6$ are newly forbidden,
and continuous 2-form symmetries in $d=6$ newly refuse anti-self-dual conserved currents.
The results are shown in Table~\ref{tab:MRsusy}, 
which only depend on the existence of the supersymmetry
but not on its amount $\CN$ in all dimensions.
\begin{table*}[h]
	\begin{align*}
	    \renewcommand{\arraystretch}{1.2}
    	\begin{array}{|c|c|c|c|c|}
        	\hline
        	& d=3
        	& d=4
        	& d=5
        	& d=6\\
        	\hline
        	p=0
        	& -
        	& -
        	& -
        	& -\\
        	p=1
        	& -
        	& -
        	& \text{\textcolor{red}{\ding{55}}}
        	& \text{\textcolor{red}{\ding{55}}}\\
        	p=2
        	&
        	& -
        	& -
        	& \text{\textcolor{red}{\ding{55}}}: \text{if ASD}\\
        	p=3
        	&
        	&
        	& -
        	& -\\
        	p=4
        	&
        	&
        	&
        	& -\\
        	\hline
    	\end{array}
	\end{align*}
	\caption{
    	Newly forbidden $p$-form symmetries under the presence of supersymmetry.
	}\label{tab:MRsusy}
\end{table*}

\newpage

\bibliographystyle{ytphys}
\baselineskip=.95\baselineskip
\bibliography{CFTnogo}

\end{document}